\documentstyle[onecolumn]{mn}

\newif\ifAMStwofonts

\ifoldfss
  \ifCUPmtlplainloaded \else
    \NewTextAlphabet{textbfit} {cmbxti10} {}
    \NewTextAlphabet{textbfss} {cmssbx10} {}
    \NewMathAlphabet{mathbfit} {cmbxti10} {} 
    \NewMathAlphabet{mathbfss} {cmssbx10} {} 
  \fi
  \ifAMStwofonts
    \ifCUPmtlplainloaded \else
      \NewSymbolFont{upmath} {eurm10}
      \NewSymbolFont{AMSa} {msam10}
      \NewMathSymbol{\upi}     {0}{upmath}{19}
      \NewMathSymbol{\umu}     {0}{upmath}{16}
      \NewMathSymbol{\upartial}{0}{upmath}{40}
      \NewMathSymbol{\leqslant}{3}{AMSa}{36}
      \NewMathSymbol{\geqslant}{3}{AMSa}{3E}

       \let\le=\leqslant
       \let\ge=\geqslant
    \fi
  \fi
\fi 

\ifnfssone
  \newmathalphabet{\mathit}
  \addtoversion{normal}{\mathit}{cmr}{m}{it}
  \addtoversion{bold}{\mathit}{cmr}{bx}{it}
  \newmathalphabet{\mathbfit} 
  \addtoversion{normal}{\mathbfit}{cmr}{bx}{it}
  \addtoversion{bold}{\mathbfit}{cmr}{bx}{it}
  \newmathalphabet{\mathbfss} 
  \addtoversion{normal}{\mathbfss}{cmss}{bx}{n}
  \addtoversion{bold}{\mathbfss}{cmss}{bx}{n}
  \ifAMStwofonts
    \ifCUPmtlplainloaded \else
      \UseAMStwoboldmath
      \makeatletter
      \new@mathgroup\upmath@group
      \define@mathgroup\mv@normal\upmath@group{eur}{m}{n}
      \define@mathgroup\mv@bold\upmath@group{eur}{b}{n}
      \edef\UPM{\hexnumber\upmath@group}
      \new@mathgroup\amsa@group
      \define@mathgroup\mv@normal\amsa@group{msa}{m}{n}
      \define@mathgroup\mv@bold\amsa@group{msa}{m}{n}
      \edef\AMSa{\hexnumber\amsa@group}
      \makeatother
      \mathchardef\upi="0\UPM19
      \mathchardef\umu="0\UPM16
      \mathchardef\upartial="0\UPM40
      \mathchardef\leqslant="3\AMSa36
      \mathchardef\geqslant="3\AMSa3E

       \let\le=\leqslant
       \let\ge=\geqslant
    \fi
  \fi
\fi 

\ifnfsstwo
  \DeclareMathAlphabet{\mathbfit}{OT1}{cmr}{bx}{it}
  \SetMathAlphabet\mathbfit{bold}{OT1}{cmr}{bx}{it}
  \DeclareMathAlphabet{\mathbfss}{OT1}{cmss}{bx}{n}
  \SetMathAlphabet\mathbfss{bold}{OT1}{cmss}{bx}{n}
  \ifAMStwofonts
    \ifCUPmtlplainloaded \else
      \DeclareSymbolFont{UPM}{U}{eur}{m}{n}
      \SetSymbolFont{UPM}{bold}{U}{eur}{b}{n}
      \DeclareSymbolFont{AMSa}{U}{msa}{m}{n}
      \DeclareMathSymbol{\upi}{0}{UPM}{"19}
      \DeclareMathSymbol{\umu}{0}{UPM}{"16}
      \DeclareMathSymbol{\upartial}{0}{UPM}{"40}
      \DeclareMathSymbol{\leqslant}{3}{AMSa}{"36}
      \DeclareMathSymbol{\geqslant}{3}{AMSa}{"3E}

       \let\le=\leqslant
       \let\ge=\geqslant
    \fi
  \fi
\fi 

\ifCUPmtlplainloaded \else
  \ifAMStwofonts \else 
    \def\upi{\pi}
    \def\umu{\mu}
    \def\upartial{\partial}
  \fi
\fi
\def\simlt{\lower.5ex\hbox{$\; \buildrel < \over \sim \;$}}
\def\simgt{\lower.5ex\hbox{$\; \buildrel > \over \sim \;$}}

\title{Dark matter and visible baryons in M33}

\author[E. Corbelli]
{Edvige Corbelli
\\
Osservatorio Astrofisico di Arcetri, Largo E. Fermi,5
I-50125 Firenze, Italy\\
\      
         }

\date{Received .... ; accepted .....}

\begin{document}

\maketitle  

\label{firstpage}

\begin{abstract}
 
In this paper we present new measurements of the gas kinematics in M33 using 
the CO J=1-0 line. The resulting rotational velocities complement previous 
21-cm line data for a very accurate and extended rotation curve of this 
nearby galaxy. The implied dark matter mass, within the total gaseous extent, 
is a factor 5 higher than the visible baryonic mass. Dark matter density profiles 
with an inner cusp as steep as $R^{-1}$, suggested by some numerical simulation 
of structures formation, are compatible with the actual data. The dark matter 
concentrations required for fitting the M33 rotation curve are very low but still 
marginally consistent with halos forming in a standard Cold Dark Matter cosmology. 
The M33 virialized dark halo is at least 50 times more massive than the visible 
baryons and its size is comparable with the M33-M31 separation. Inner cusps as 
steep as $R^{-1.5}$ are ruled out, while halo models with a large size core of 
constant density are consistent with the M33 data. A central spheroid of stars 
is needed and we evaluate its dynamical mass range. Using accurate rotational 
velocity gradients and the azimuthally averaged baryonic surface densities, we 
show that a disk instability can regulate the star formation activity in M33. 
Considering the gaseous surface density alone, the predicted outer star formation 
threshold radius is consistent with the observed drop of the H-$\alpha$ surface 
brightness if a shear rate criterion is used with the lowest possible value of 
velocity dispersion. The classical Toomre criterion predicts correctly the size 
of the unstable region only when the stellar or dark halo gravity, derived in 
this paper, is added to that of the gaseous disk.

\end{abstract}


\vfill
\eject

\section{Introduction}

M33 is a low luminosity spiral galaxy in the Local Group. Its large angular
extent and well determined distance make it ideal for a detailed study
of the radial distribution of visible and dark matter. The interplay
between the dark and visible matter as well as the nature and
the radial distribution of dark matter are still open crucial issues with 
strong implications for galaxy formation and evolution theories. M33 has 
three advantages over its brighter neighbor M31: {\it{(i)}} being of lower 
luminosity, it is a dark matter dominated spiral in which it is easier to 
disentangle the dark and luminous mass components (Persic, Salucci $\&$
Stel 1996, Corbelli $\&$ Salucci 2000 hereafter CS); {\it{(ii)}} it has 
no prominent black hole or bulge component in the center (Gebhardt et al. 2001) 
and therefore it can be used to test if the innermost dark matter density 
profile is consistent with cosmological models which predict a central density 
cusp; {\it{(iii)}} being at lower declination, high sensitivity 21-cm 
observations with the Arecibo radio telescope made possible a velocity field 
map out to large galactocentric radii. 
CS have used this velocity field map, together with aperture synthesis data
of Newton (1980) and with the tilted ring model results
of Corbelli $\&$ Schneider (1997), to derive the rotation curve of M33 out to 13 
disk scalelengths. This unveiled large scale properties of the mass
distribution in M33, such as the lack of correlation between the dark and visible
matter, the dark halo mass and the possible radial slope of the dark matter 
density. However some ambiguities are left, especially in the inner regions, 
where the low spatial resolution of 21-cm observations and the possible role of
a stellar nucleus or of the molecular gas component have not allowed to establish 
unambiguously the steepness of the dark matter density profile.

Several authors, e.g. Bolatto et al. (2002), have shown the advantages of 
using CO observations for determining the central densities of dark matter halos.
A full map of the CO emission in M33 would not only provide 
higher spatial resolution data on the velocity field but it would also 
give a first estimate of the molecular mass distribution in this galaxy.
The very blue color of M33 and the presence of prominent HII regions indicate  
vigorous star formation activity in the nuclear region and along the spiral
arms: here the molecular gas component might represent a substantial fraction
of the total disk mass. CO observations of the innermost 500 pc of M33
(Wilson $\&$ Scoville 1989), have confirmed in fact that here the molecular gas 
has a higher surface density than the atomic gas. Recently Heyer, Corbelli 
$\&$ Schneider (2003) have fully mapped the CO emission in M33 using the 
FCRAO 14-m telescope. A detailed analysis of the molecular gas distribution and 
emission features will be presented elsewhere. In this paper we derive only the 
average radial distribution of the molecular gas mass (Section 2) and analyze  
the kinematics of the CO gas (Section 3). The implications for the dark matter 
halo models are shown in Section 4. An accurate knowledge of the velocity field 
and of the mass surface densities over a large disk area are useful for 
understanding the processes which trigger the star formation in this galaxy, 
such as disk instabilities, spiral arms, cooling and heating mechanisms. A possible
link between the radial extent of the observed star formation region and the
gravitationally unstable region in the disk is discussed in Section 5. 

\section{The molecular data}

The nearby spiral galaxy M33 has been fully mapped in the CO J=1-0 transition  
with a velocity resolution of 0.81 km s$^{-1}$ and a spatial resolution 
of 45 arcsec using the FCRAO 14-m telescope (Heyer, Corbelli $\&$ Schneider 
2003). A 22.5 arcsec spatial sampling is obtained over an area of about 900 
arcmin$^2$ centered on RA 01$^h$33$^m$50$^s$.89 and DEC 30$^\circ$39'36".7.  
For a distance to M33 of 0.84 Mpc (Freedman et al. 2001), which will 
be used throughout this paper, this means that the total disk area
observed is about 44 kpc$^2$ at a spatial resolution of 165 pc.
Here we shall use the data at the original
spatial and spectral resolution to derive the kinematical information on
the molecular gas. We analyze in detail each spectrum by assigning 
a spectral window $\Delta V = V^f-V^i$ to any detectable emission feature. 
There are a few cases where more than one spectral window is needed but 
secondary signals are always quite low. If no emission is clearly present 
in a given spectrum, a spectral window is set based on the diffuse emission 
detectable when averaging nearby positions. 
We compute the integrated brightness temperature $F$ in K km s$^{-1}$ by 
integrating over $\Delta V$ the telescope temperature observed at each 
position in the sky, and dividing this number by the main beam 
efficiency of the telescope (which is 0.45 at 115 GHz). 
In order to derive the average radial distribution of the molecular gas
we need a concentric ring model. 45 concentric rings of width 150 pc cover 
all the M33 surface area observed. They are tilted as to best
fit the neutral hydrogen distribution (Corbelli $\&$ Schneider 1997).
Inside this area the position angle of the best fitting ring model
changes only slightly, and therefore we consider it constant and equal 
to 22$^o$.  The inclination of the rings increases from 50$^o$ to 58$^o$ 
moving radially outwards, as described in details in Section 3 of CS.     
We use the concentric ring model for assigning a face on 
radial distance to each observed position, and for deriving
the inclination corrected average flux in each ring.
Following Wilson's results (Wilson 1995) we use the standard conversion 
factor between the CO luminosity and 
the line of sight column density of $H_2$ ($S_{H_2}=2.8\times 
10^{20}\times F$~cm$^{-2}$ = 4.5$\times F$ M$_\odot$ pc$^{-2}$).

\begin{figure}
 
\vspace{12.5cm}
\includegraphics{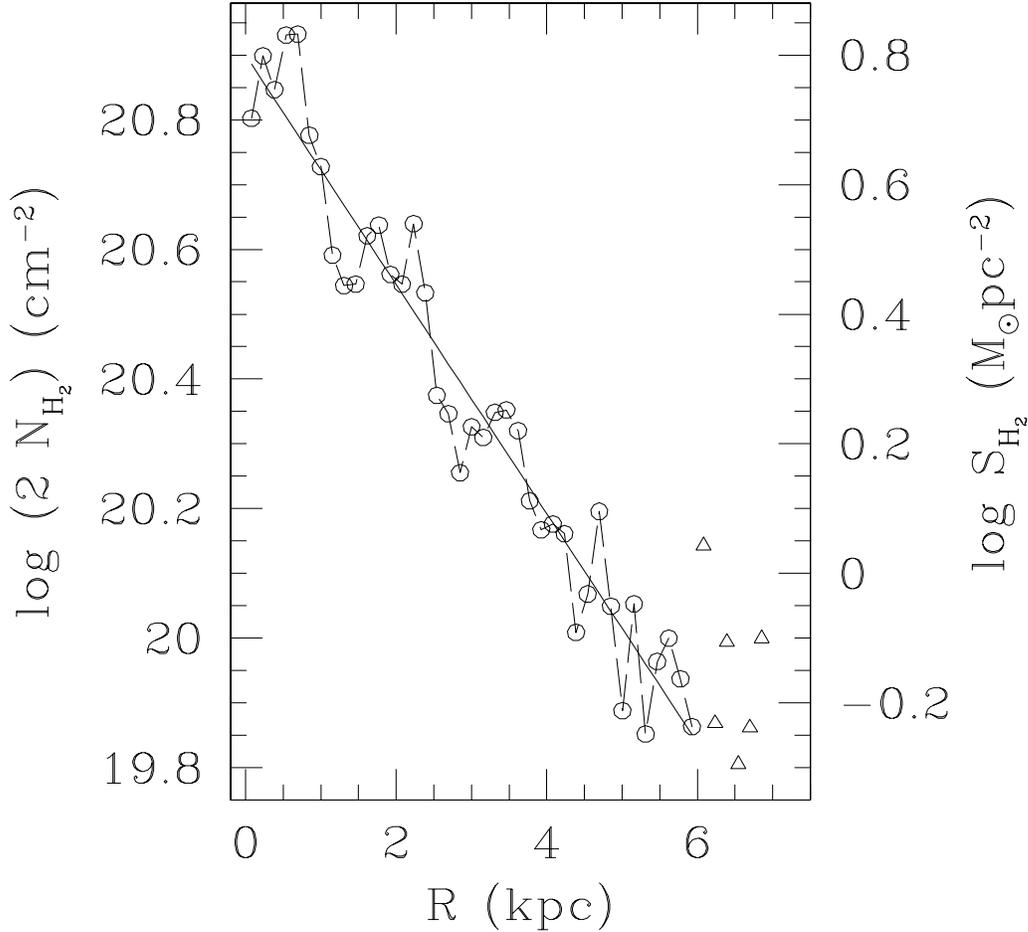}
\caption{ Average surface density of molecular hydrogen as a function of
$R$ in the plane of the M33 disk. The straight
line is the best fit to all data with $R<6$~kpc (open dots). Larger radii 
were only partially covered by our survey. }
\label{fig:fig1 }
\end{figure}

The resulting average surface density of molecular hydrogen, perpendicular 
to the plane of the galaxy (i.e. corrected for the disk inclination respect to
our line of sight), is shown in Figure 1. The estimated total mass 
of molecular gas is $2\times 10^8$ M$_\odot$. This is a factor 6 higher than 
measured by Wilson $\&$ Scoville (1989) in the smaller nuclear region. The  
molecular hydrogen mass is about 10$\%$ the neutral hydrogen mass 
($2.2\times 10^9$ M$_\odot$ for D=0.84 Mpc). The 
surface density distribution does not peak at the center but shows a peak at
a few hundred parsecs from the nucleus, as noted previously by Wilson $\&$ 
Scoville (1989). The most prominent CO emission features are located in the 
innermost 1~kpc region and along the spiral arms. This can be seen already in 
the 1.3~$\times$~7~kpc$^2$ preliminary emission map (Corbelli 2000) oriented 
along the M33 major axis. An exponential is fitted to the azimuthally averaged 
surface density data from 0 to 6~kpc. We leave out from the fit the last 
6 points shown in Figure 1 which seem to indicate a plateau at the level of 0.6 
M$_\odot$~pc$^{-2}$. In reality we don't have enough sensitivity and 
coverage in our survey to determine the slope of the $H_2$ distribution so far 
out. The best fit to S$_{H_2}$ is shown by the straight line in Figure 1 and
is given by

\begin{equation}
S_{H_2}= 6.3 \times {\hbox{exp}}(-{R\over2.5{\hbox{ kpc}}})\ 
\ {\hbox{M}}_{\odot}{\hbox{pc}}^{-2}
\end{equation}

\begin{figure}
 
\vspace{12.5cm}
\includegraphics{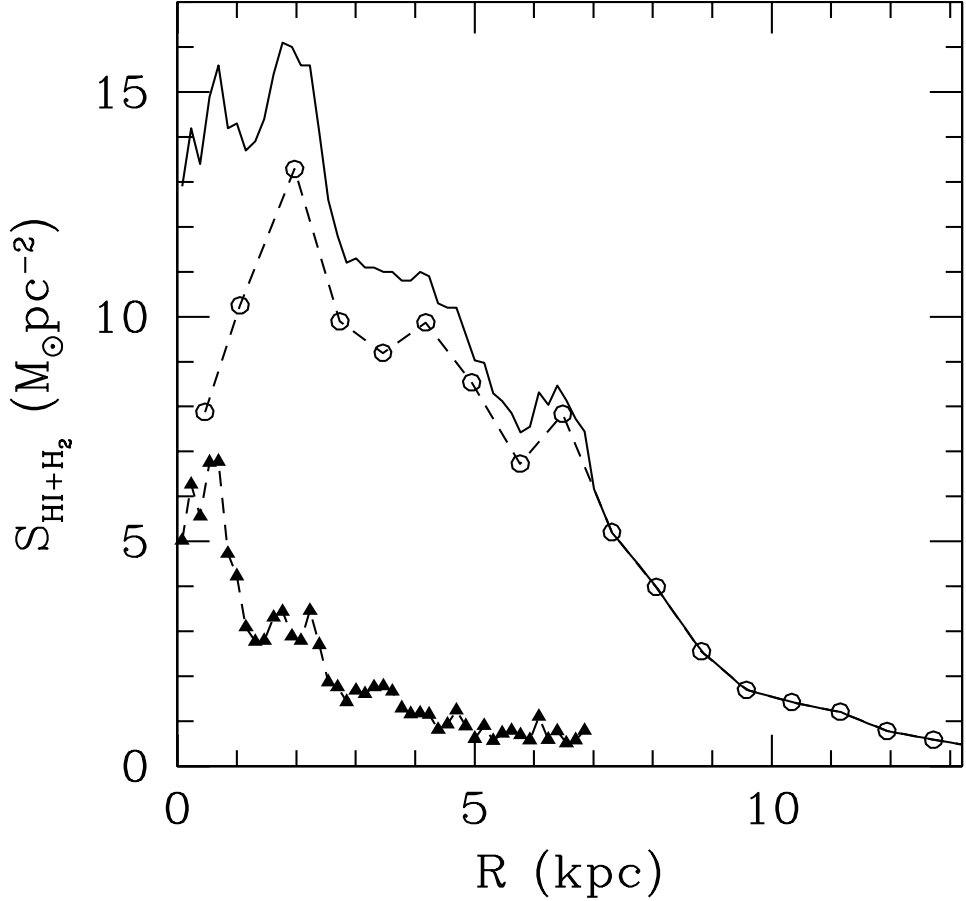}
\caption{ Average surface density of molecular hydrogen (filled triangles)
and of atomic hydrogen (open circles) as a function of $R$ in the plane of 
the M33 disk. The solid line is the sum of the two surface densities. }
\label{fig:fig2 }
\end{figure}

Therefore the CO radial scalelength in  M33 is larger than the B and K band 
scalelengths (1.9 and 1.3 kpc respectively, Regan $\&$ Vogel 1994). In Figure 2 
we show the HI as well as the total (HI + H$_2$) gas distribution in 
this galaxy. The total gas surface density  can be described by a 
gaussian profile with half width at half maximum of about 6~kpc. 
Note that the bump of the spiral arm feature around 2~kpc is visible in the 
azimuthal averages of the molecular and atomic gas shown in Figure 2.

\section{The CO rotation curve}

As Wilson $\&$ Scoville (1989) have already pointed out from a comparison of
interferometric and single dish data, small clouds and diffuse emission in
M33 contain a non negligible fraction of the total molecular mass.
Half of the molecular gas mass that we estimate in M33 resides in fact in
regions with integrated flux less than 3$F_\sigma \simeq$ 1.33 K km s$^{-1}$.
$F_\sigma $ is the integrated noise defined in terms of the spectral window
$\Delta V$ as:

\begin{equation}
F_{\sigma}=\sigma_{rms}\times \sqrt{\Delta V \times 0.81}
\end{equation}

\noindent
where $\sigma_{rms}$ is the noise in the spectrum at the original 
resolution of 0.81 km s$^{-1}$.  We fit single or double gaussians to 
all signals with integrated flux $F>F_{5\sigma}$ with a success rate of 75$\%$. 
There is no significant shift between the gaussian and the flux weighted mean 
velocities (difference is $\simlt 2$ km s$^{-1}$ and is never above 5 km 
s$^{-1}$). In deriving kinematical informations we shall consider only spectra 
with $F>5F_{\sigma}$, well fitted by gaussian profiles, from positions at angles
$\alpha < 45^\circ$ from the major axis. These conditions restrict the analysis 
to a smaller radial range than observed. 
Luckily the prominent CO features come from locations close to the major axis 
and therefore we do have many points to average for building up a rotation curve.
Corrections for the inclination of the disk and for the angle $\alpha$ between 
the spectrum position and the major axis are applied according to the tilted 
ring model described in the previous Section. 
Radial bin averages of the flux weighted mean velocities are plotted in Figure 3,
also for the northern and southern side of the galaxy separately. 
Error bars in each bin show the dispersion of the velocities about the mean. 

\begin{figure}
 
\vspace{12.5cm}
\includegraphics{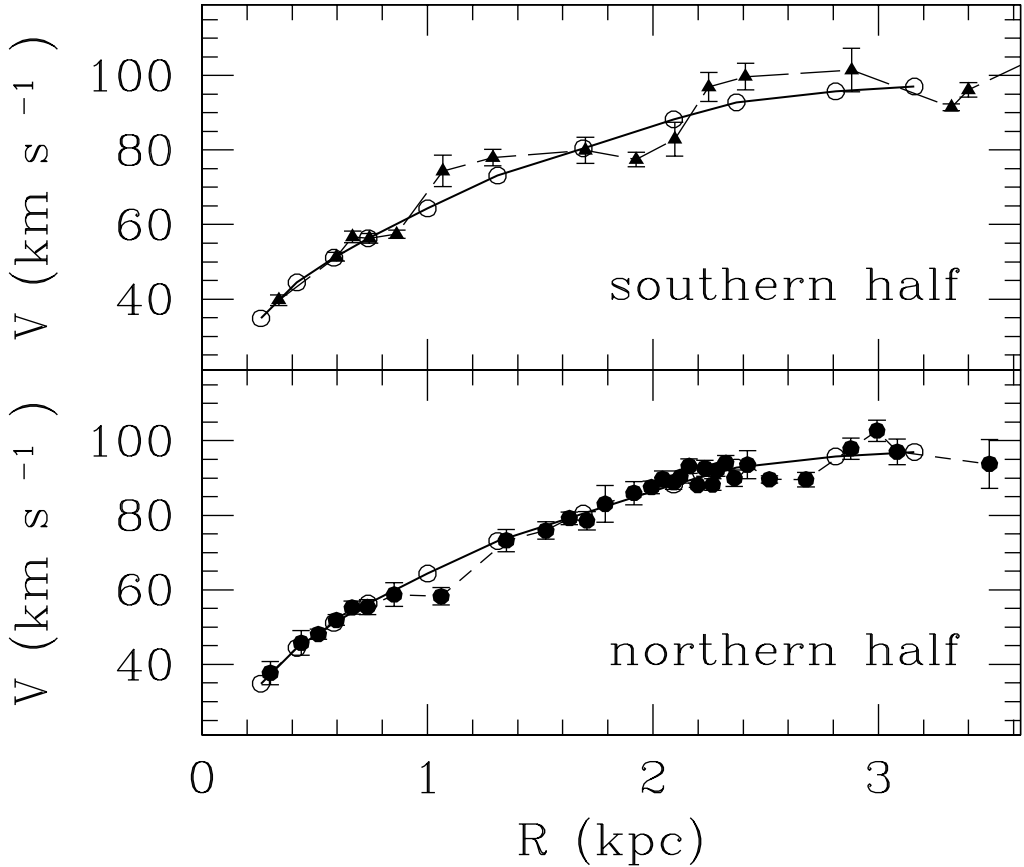}
\caption{The northern and southern half data after averaging 
10 adjacent points in each radial bin.  Error bars show the dispersion  
in each bin. The rotation curve derived from averaging the
northern and southern data is shown with open circles
connected by a heavy solid line. }
\label{fig:fig3 }
\end{figure}

In the innermost 1~kpc the south and north rotation curves coincide
reasonably well. Beyond this radius there are some differences both in
the dispersions and in the mean values of $V$. Velocities in the southern 
half of M33 have higher dispersion than in the northern half. Also the 
southern side shows more clearly wiggles which can be interpreted as signs of 
density wave perturbations associated with the spiral arms. The detection of 
such features in the northern side is more doubtful. Residual velocities, 
obtained by subtracting the average rotational velocity $V(R)$ from the 
rotational velocity measured in each spectrum, $V_{obs}(R)$, are shown in 
Figure 4. In the left hand panel of Figure 4 the map of residual velocities is 
superimposed to the used tilted ring model, after rotating the galaxy by 22$^o$ 
clockwise in the plane of the sky. Cross and square symbols are for positive
and negative residuals respectively, and their size is proportional to the
amplitude of the residual velocities. Since the approaching side of the
galaxy is the northern side ($y>0$), a positive residual there means that
the local velocity is less than the average rotational velocity. For each 
quadrant of the galaxy, the right hand panels of Figure 4
show the residual velocities as function of the galactocentric radius. 
It hard to judge whether there are signs of non circular motions and 
spiral density waves in the velocity residuals map. Part of the difficulty is
due to the fact the bulk of the CO emission comes from localized regions and 
is not smoothly distributed throughout the disk. A better attempt can be made 
in the future with the use of a numerical simulation for any given model 
assumption.

\begin{figure}
 
\vspace{12.5cm}
\includegraphics{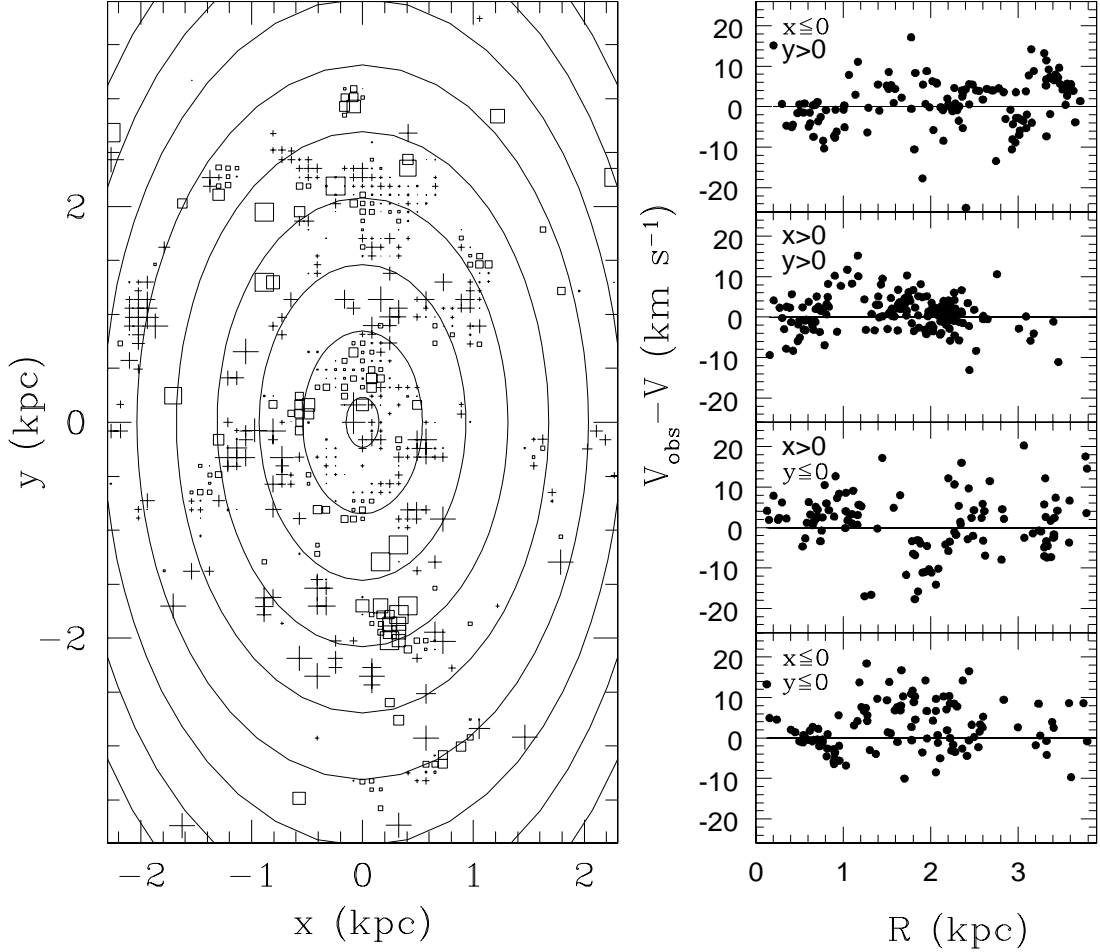}
\caption{The left hand panel shows the map of the residual velocities,
$V_{obs}-V$, where $V$ is the average rotational velocity at $R$. It is  
superimposed to the used tilted ring model, after rotating the galaxy by 22$^o$ 
clockwise in the plane of the sky. Cross and square symbols are for positive
and negative residuals respectively. The size of the symbols is proportional to 
the amplitude of the residuals. In the right hand panels of Figure 4 
the residual velocities are shown as function of galactocentric radius 
in each quadrant of the galaxy.}
\label{fig:fig4 }
\end{figure}

Comparison of the rotation curve in Figure 3 with the 21-cm data of Newton 
(1980) enlightens a discrepancy in the northern inner region, where at
$R\simlt 0.5$~kpc the HI velocities are lower by about 10 km s$^{-1}$ than 
the CO velocities. This might be due to the lack of atomic gas in the innermost 
region, which together with the larger 21-cm beam, limits the accuracy of the 
atomic line measure. Finite angular resolution is not affecting significantly 
the velocities measured in our CO survey. To estimate any possible ``beam smearing'' 
effect we assume that the true velocity, for a given mass model, is that  
resulting from rotation curve and perform a numerical simulation to derive the 
observed velocities (van den Bosch $\&$ Swaters 2001). The observed velocities at 
each position on the plane of the sky are the flux weighted means of the rotational
velocities along the line of sight, corrected for the assumed ring model and 
convolved with the shape of the telescope beam. We then compare the binned data 
with equally binned values of the velocities resulting from the simulation (we 
exclude points which form an angle $\alpha>45^\circ$ with the major axis).
Differences between the observed and the true velocities, for mass
models which give a good fit to the data, are always smaller than 0.6 km s$^{-1}$ 
except in the innermost 0.5~kpc where the beam smearing effect can lower the
observed velocity as much as 2 km s$^{-1}$.

\section{The mass components and constrains on cosmological halo models}

Rotational velocities, which we use to derive the visible and dark 
matter distributions in M33, are shown in Figure 5. The innermost point at
$R\simeq 150$~pc is from H$\alpha$ measurements of Rubin $\&$ Ford (1985). Our  
CO data extending over the interval $0.2\le R<3.4$~kpc have been averaged into 
the subsequent 11 bins. For $3.4\le R< 5.5$~kpc data are taken from 21-cm 
observations of Newton (1980) at the Cambridge Half-Mile telescope while for 
larger radii we use Arecibo 21-cm data (CS and references therein). The plotted 
error bars in each bin are twice the dispersion of the velocities about the mean.

\begin{figure}
 
\vspace{15.cm}
\includegraphics{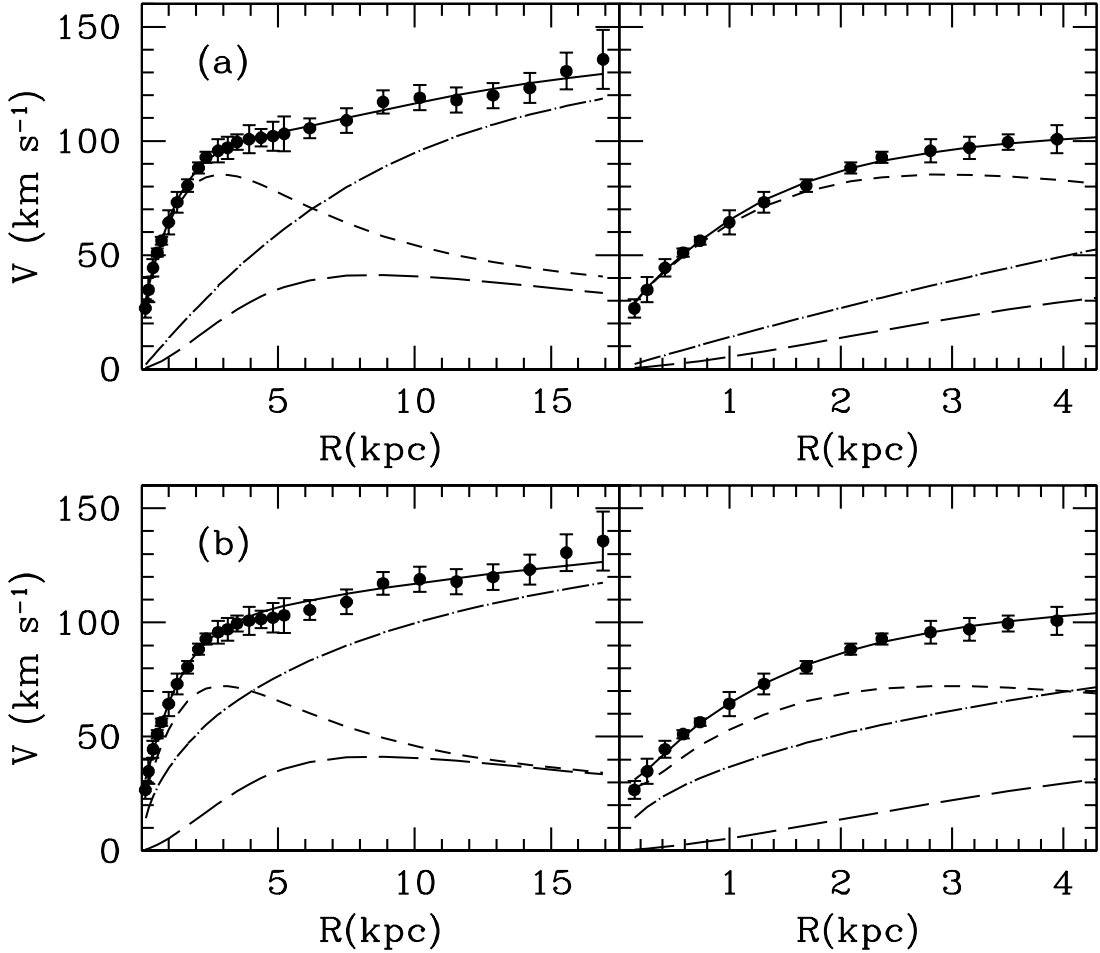}
\caption{The M33 rotation curve (points). $(a)$The best-fitting
model (solid line) using Burkert profile for the dark halo 
density distribution, and the nuclear model (1). Also shown are 
the dark halo contribution (dot-dashed line), the stellar disk 
+ nucleus (short-dashed line) and the gas contribution 
(long-dashed line). $(b)$The same as in $(a)$ but for NFW dark 
halo profile with $C_{100}=8$ and the nuclear model (2).
The left panels show an enlargement of the inner fit.} 
\label{fig:fig5 }
\end{figure}

We consider the dynamical contribution of two visible mass components: 
stars, distributed in a disk and in the nucleus, and gas, in molecular 
and in neutral atomic form. For the stellar disk we consider that
the mass follows the light distribution in the K-band, well fitted by  
an exponential with scalelength $R_d\sim 1.3 \pm 0.2$~kpc (Regan $\&$ Vogel 
1994). After subtracting the exponential disk from the surface brightness, 
a central emission excess remains at $R\simlt 0.5$~kpc (Bothun 1992, 
Regan $\&$ Vogel 1994) which has been attributed to spiral arms, to
an extended semistellar nucleus or to a diffuse halo or bulge. 
We shall refer to this excess as the ``nucleus'' of M33 despite the fact 
that in the literature the word ``nucleus'' is often used for the light 
excess observed only at $R \simlt 10$~pc (Kormendy $\&$ McClure 1993).  
Nuclear rotational velocities have been observed in H$\alpha$ along the 
major axis between 20 and 200~pc from the center (Rubin $\&$ Ford 1985), 
and seem roughly constant at about $\sim 27$~km s$^{-1}$. 
This flatness of the rotation curve is quite puzzling and
needs further investigation once uncertainties on the nuclear rotational 
velocities will be established together with a knowledge of the velocity
field along different directions. In this paper we model the rotation curve
at $R\simgt 150$~pc only, considering two different possibilities
for the nuclear component $V_n$ at these radii:

\noindent
$(1)$ $V_{n}\propto R^{0.5}/(R+a)$. This rotational velocity approximates  
the dynamical contribution of a spheroidal mass distribution whose surface 
density can be fitted by a de Vaucouleurs $R^{1/4}$ law (Hernquist 1990). 
This scaling law has been used by Regan $\&$ Vogel (1994) to fit the J-band
nuclear surface brightness. The compatibility of the fit   
with photometric data in other bands (e.g. Bothun 1992) is somewhat 
uncertain, as is the value of the scalelength $a$. Suggested values of $a$ 
lie in the range $0.1\le a\le 1$~kpc, which we will consider in this paper.
  
\noindent
$(2)$ $V_{n}\propto R^{-0.115}$, as if core collapse has happened    
or is happening in the compact stellar nucleus (Cohn 1980, Kormendy $\&$
McClure 1993). The corresponding power-law mass density distribution 
develops up to a limiting outer radius $R_c$, which should also be given.
For $R>R_c$, $V_{n}\propto R^{-0.5}$.
We consider as possible $R_c$ values the radius of each data point 
of the rotation curve inside the optical region i.e. $R_c < 8$~kpc.
We consider also the case $R_c\ll 150$~pc, which is
appropriate if most of the nuclear mass resides at very small radii and 
the nucleus can be treated as a point source throughout the extent of 
the present rotation curve.

Hereafter, we will refer to these models as nuclear model (1) and (2)
respectively. It has been established by CS that the visible baryons in 
M33 are only a small fraction of the total galaxy mass, and that a dark 
matter halo is in place. We will use the M33 rotation curve presented here 
to test in details the consistency of the required halo density profile 
with theoretical models which predict a well defined dark matter 
distribution around galaxies. Namely, in addition 
to the widely used non-singular isothermal dark halo model we
shall consider a spherical halo with a dark matter density profile as
originally derived by Navarro Frenk and White (1996, 1997, hereafter NFW) 
for galaxies forming in a Cold Dark Matter scenario (hereafter CDM). We
consider also the Burkert and the Moore dark matter density profile 
(Burkert 1995, Moore et al. 1998). All models have two free parameters
to be determined from the rotation curve fit. The density distribution for 
a non-singular isothermal dark halo can be written as:

\begin{equation}
\rho(R)= {\rho_{iso}\over 1+({R\over R_{iso}})^2}
\end{equation}

\noindent
The density profile proposed by Burkert (1995) has also a flat density core:

\begin{equation}
\rho(R)={\rho_B\over (1+{R\over R_B})\Bigl(1+({R\over R_B})^2\Bigr)}
\end{equation}

\noindent
and a strong correlation between the two free parameters $\rho_B$ and $R_B$ has
been found by fitting the rotation curve of low mass disk galaxies.

\noindent
The NFW density profile is:

\begin{equation}
\rho(R)={\rho_{NFW} \over {R\over R_{NFW}}\Bigl(1+{R\over R_{NFW}}\Bigr)^2}
\end{equation}

\noindent
Numerical simulations of galaxy formation in CDM models find a 
correlation between $\rho_{NFW}$ and $R_{NFW}$ which depends on the 
cosmological model (Avila-Reese et al. 2001, Eke Navarro and Steinmetz 2001). 
Often this correlation is expressed using the concentration parameter 
$C_{\Delta}\equiv R_{\Delta}/R_{NFW}$ and $M_{\Delta}$ or $V_{\Delta}$. 
$R_{\Delta}$ is the radius of a sphere containing a mean density $\Delta$ 
times the critical density, $V_{\Delta}$ and $M_{\Delta}$ are the 
characteristic velocity and mass inside $R_{\Delta}$. We use a Hubble constant
$H_0=65$~km s$^{-1}$ Mpc$^{-1}$ and show results for $\Delta=100$, 
as in CDM halo models of Eke Navarro and Steinmetz (2001), unless otherwise 
specified. 

\noindent
A similar profile but with a different central slope has been suggested 
instead by numerical simulations of Moore et al. (1998) for galaxies 
forming in CDM scenarios:  

\begin{equation}
\rho(R)={\rho_M \over ({R\over R_M})^{1.5} \Bigl(1+({R\over R_M})^{1.5}\Bigr)}
\end{equation}

For each dark halo model and each nuclear model we shall determine five
free parameters applying the least-square method to the observed rotation curve. 
These free parameters are: the dark halo core radius $R_{iso,NFW,M,B}$, the
dark matter density $\rho_{opt}$ computed at $R=R_{opt}\equiv 3.2R_d\simeq 
4.1$~kpc, the ratio between the stellar disk mass $M_d$ and the blue luminosity 
of the galaxy $L$ ($L\simeq 5.7\times 10^9$~L$_\odot$), the nuclear 
mass $M_n$, and the nuclear scale length $a$ or the cut-off radius $R_c$. 
The 1-$\sigma$, 2-$\sigma$ and 3-$\sigma$ ranges are determined  using the 
reduced $\chi^2$ and assuming Gaussian statistics. They are computed by 
determining in the free parameters space the projection ranges, along each 
axis, of the hypersurfaces corresponding to the 68.3$\%$, 95.4$\%$ and 
99.7$\%$ confidence levels.

Both the non-singular isothermal dark halo model and the Burkert dark 
matter density profile give good fits to the rotation curve since the
minimum reduced $\chi^2$ is  0.7 for the nuclear model (1) and 1.2 for the
nuclear model (2). The best fitting values of the free parameters
are: $a=0.4$~kpc, $M_n=3\times 10^8$~M$_\odot$,
$M_d/L=1~$M$_\odot$/L$_\odot$, $R_{iso}=7$~kpc or $R_B=12$~kpc, 
$\rho_{opt}=5\times 10^{-25}$~g cm$^{-3}$ for the nuclear model (1), 
and $R_c=0.7$~kpc, $M_n=10^8$~M$_\odot$,
$M_d/L=1$~M$_\odot$/L$_\odot$, $R_{iso}=8$~kpc or $R_B=13$~kpc, 
$\rho_{opt}=5\times 10^{-25}$~g cm$^{-3}$ for the nuclear model 
(2). The 3-$\sigma$ ranges for the free parameters 
of both the isothermal and the Burkert model are:

$$0.8\le {{\hbox{M}}_d/{\hbox{L}}\over {\hbox{M}}_\odot/{\hbox{L}}_\odot}
 \le 1.1; \qquad 5\times 10^7\le {{\hbox{M}}_n\over {\hbox{M}}_\odot}\le 8
\times 10^8; \qquad 4\times 10^{-25}\le {\rho_{opt}\over {\hbox{g cm}}^{-3}}
\le7\times 10^{-25};$$

$$0.4\le{R_c \over {\hbox{kpc}}}<8; \qquad 7\le{R_B\over {\hbox{kpc}}}\le 
19\qquad {\hbox{or}}\qquad 4\le {R_{iso}\over {\hbox{kpc}}}\le 13.$$

The resulting 3-$\sigma$ range for the nuclear scale length $a$ in
the nuclear model (1) covers the whole range we have considered i.e.
$0.1\le a\le 1$~kpc. For the nuclear model (2) instead we find a 
lower limit to $R_c$. This means that the nucleus cannot be  
considered a point source for any possible dark halo models we 
are examining. This `dynamical' conclusion agrees with photometric 
data (Bothun 1992, Minniti, Olszewski $\&$ Rieke 1993, 
Regan $\&$ Vogel 1994) showing an excess of light inward of 0.5~kpc 
after subtracting the exponential disk.
Unfortunately estimates of the nuclear luminosity $L_n$ are still 
quite uncertain: in the B-band $7\times 10^7\le L_{n,B}\le 3\times 
10^8$~L$_{\odot,B}$ (Bothun 1992, Regan $\&$ Vogel 1994), and we cannot
limit the nuclear stellar mass range further based on possible
unrealistic values of the mass to light ratio (Bell $\&$ de Jong 2001). 
An extended nucleus is not required instead if the density of the 
dark matter scales as $R^{-1.3\pm 0.1}$ throughout the halo;
these type of solutions has been investigated by CS and provides
also good fits to the rotation data presented in this paper. 
Both the Burkert and the isothermal model require a very extended
core region of nearly constant dark matter density. Figure 5$(a)$ 
shows the best fit using the Burkert profile with the nuclear model 
(1). The 1-$\sigma$ and 3-$\sigma$ probability contours for $a=0.4$~kpc
in the $\rho_{opt}-R_B$ plane are displayed in Figure 6$(a)$. The 
correlation found by Burkert between $\rho_B$ and $R_B$ implies a 
relation between $\rho_{opt}$ and $R_B$ shown with crosses in Figure 
6$(a)$. We can see that the M33 acceptable parameters lie close to, 
but outside, the Burkert correlation line. So either the found scaling 
relation between $\rho_{opt}$ and $R_B$ should be slightly modified to 
include M33, or it changes for values of $R_B$ larger than those 
considered by Burkert.
A Burkert model which fits M33 is indistinguishable from an 
isothermal model since both require a large constant density core 
region and, even at the outermost fitted radius, the dark matter 
density does not decline yet as an $R^{-2}$ or $R^{-3}$ power law.
The dark matter density profile suggested by Moore fails
to fit the observed rotation curve in M33.
 
\begin{figure}
 
\vspace{8.cm}
\includegraphics{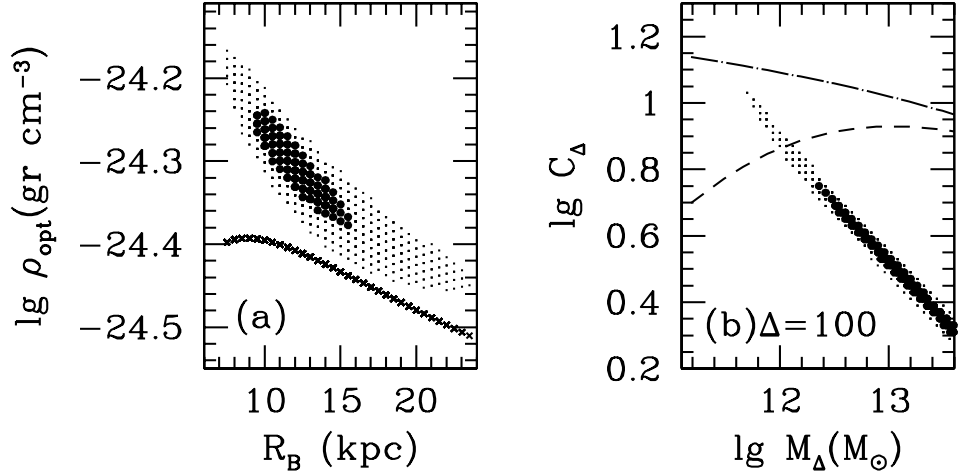}
\caption{$(a)$The light dots indicate the 3-$\sigma$  
confidence area for the Burkert dark halo model in the $R_B$
lg($\rho_{opt}$) plane; the heavy dots are for the 1-$\sigma$ 
confidence area. The cross line indicates the correlation
found by Burkert. The 3-$\sigma$ (light dots) and 1-$\sigma$ 
(heavy dots) confidence areas for the NFW dark halo model are 
shown in $(b)$ in the lg($M_\Delta$)-lg($C_\Delta$) plane for
$\Delta$=100. The dashed and dashed-dotted lines indicates  
the numerical simulation results for model $W_8$ and
$S_{0.9}$ respectively of Eke, Navarro $\&$ Steinmetz (2001).}
\label{fig:fig6 }
\end{figure}

For the NFW dark matter density profile for $\Delta=100$ or 200
the minimum $\chi^2$ is 0.7 for the nuclear model (1) and 1.0 for 
the nuclear model (2). The best fitting values of the free parameters  
for the nuclear model (1) are: 
$a=0.4$~kpc, $M_n=2\times 10^8$~M$_\odot$, $M_d/L=0.8$~M$_\odot$/L$_\odot$, 
$R_{NFW}=140$~kpc, $\rho_{opt}=6\times 10^{-25}$~g~cm$^{-3}$. 
For the nuclear model (2) $R_c=0.7$~kpc, $M_n=8\times 10^7$~M$_\odot$,
$M_d/L=0.8$~M$_\odot$/L$_\odot$, $R_{NFW}=300$~kpc, 
$\rho_{opt}=6\times 10^{-25}$~g~cm$^{-3}$. The 3-$\sigma$ ranges for 
the free parameters are:

$$0.5\le {{\hbox{M}}_d/{\hbox{L}}\over {\hbox{M}}_\odot/{\hbox{L}}_\odot}
 \le 0.9; \qquad 3\times 10^7\le {{\hbox{M}}_n\over {\hbox{M}}_\odot}\le 
4\times 10^8; \qquad 5\times 10^{-25}\le {\rho_{opt}\over {\hbox{g cm}}^{-3}}
\le8\times 10^{-25};$$

$$0.3\le{R_c \over {\hbox{kpc}}}<8; \qquad {R_{NFW}\over 
{\hbox{kpc}}}> 15$$

The corresponding 3-$\sigma$ upper limit for the concentrations is 
$C_{100}< 12.5$
for nuclear model (1) and $C_{100}< 11$ for nuclear model (2),
while the corresponding best fits to the rotation curve require 
$C_{100}\simeq 4$  and $C_{100}\simeq 2.5$ respectively.
For all the dark halos we are examining the nuclear model (2) provides 
a flatter rotation curve than model (1) inward of the center-most data 
point fitted, in agreement with the preliminary measures of Rubin $\&$ 
Ford (1985). The central steepness of the NFW dark matter density profile 
lowers the nuclear stellar mass required to fit the rotational data.
Figure 6$(b)$ shows the 1-$\sigma$ and 3-$\sigma$ areas in the 
$M_{100}-C_{100}$ plane. The dashed and dashed-dotted lines in the same
figure are the fits to numerical simulation results for structures
formation in the cosmological models $W_8$ and $S_{0.9}$ respectively of 
Eke, Navarro $\&$ Steinmetz (2001). Both model assume matter density 
parameters $\Omega_0=0.3$ and $\Lambda_0=0.7$, a power spectrum in the
form given by Bardeen et al. (1986) with $\sigma_8$=0.9 and shape parameter 
$\Gamma=0.2$. $S_{0.9}$ corresponds to this `standard' $\Lambda$CDM 
spectrum, while $W_8$ mimics a Warm Dark Matter power spectrum because 
in addition its power is
reduced on scales smaller than that of the characteristic free streaming 
mass $M_f=8\times 10^{11}$~M$_\odot$. It can be easily seen in Figure 
$6(b)$ that the possible $C_\Delta$ and $M_\Delta$ values for M33
are consistent with a Wark Dark Matter cosmology but lie below the 
relation found in simulated standard CDM halos. These 
$M_{\Delta}-C_{\Delta}$ relations have however an associated scatter. Even 
a small scatter, $\sigma$(lg $C_\Delta$)=0.1 as suggested for galaxies which 
did not undergo major mergers (Wechsler et al. 2002), makes the M33 dark 
halo still compatible with standard $\Lambda$CDM galaxy formation models. 
We display in Figure 5$(b)$ the fit to the M33 rotation curve using a
NFW dark halo profile with $C_{100}=8$, compatible with a standard 
$\Lambda$CDM and a Wark Dark Matter cosmology. The nuclear model (2) 
is used. The reduced $\chi^2$ and the values of the free parameters 
found from the best fit are: $\chi^2=1.5$, $R_c=0.7$~kpc, $M_n=7.5\times 
10^7$~M$_\odot$, $M_d/L=0.7$~M$_\odot$/L$_\odot$, $R_{NFW}=35$~kpc,
$\rho_{opt}=7\times 10^{-25}$~g cm$^{-3}$. 

\section{Surface densities of visible matter and the global stability 
of the disk}

In the previous Section we have determined the dark matter and stellar
contribution to the M33 rotation curve. Despite the fact that the radial
scaling law of the dark matter density required for fitting the rotation 
curve is not unique, the total baryonic and dark mass up to the last 
measurable point (R=17 kpc) are determined with small uncertainties.
The total stellar mass in the M33 disk is estimated to be 3-6$\times$
10$^9$ M$_\odot$. The mass of the dark halo at the outermost observable radius 
is about a factor 10 higher than the stellar disk mass. The neutral hydrogen 
plus the molecular gas mass is  2.4$\times 10^9$ M$_\odot$.
We don't have a firm estimate of the nuclear stellar mass $M_n$, but all 
three possible dark matter models discussed in this paper require an 
extended distribution. Its mass lies in the range  $5\times 10^7\le M_n 
\le 8\times 10^8$~M$_\odot$ if the dark halo has a large size core of
constant density, while $3\times 10^7\le M_n \le 4\times 10^8$~M$_\odot$ 
if the dark halo has a central cusp as in NFW model. 
We can estimate the total surface density of visible baryons in M33 neglecting
at the moment the small contribution from the ionized gas. Figure 7 shows
the total baryonic surface density computed by adding the helium 
to the molecular and the neutral atomic gas (a 1.33 factor correction), 
and adding to this the nuclear and stellar disk surface density. We use
the mass model displayed in Figure 5$(b)$, for which we show also the 
dark matter surface density within 0.5~kpc from the plane of the galaxy.
We extrapolate the fits of the stellar and molecular gas surface density
out to 16~kpc but it is likely they experience a truncation at smaller radii.  
For the HI surface density only we know that it declines sharply beyond 
15-16 kpc due to ionization effects by an extragalactic background (Corbelli 
$\&$ Salpeter 1993). The gas distribution is likely to extend further out but 
the dominant phase becomes ionized atomic hydrogen.  
Assuming a gaseous vertical extent of 0.5~kpc above and below the galactic 
plane, we can say that the surface density of visible baryons falls 
below the dark matter surface density contained in the M33 disk for 
$R\simgt 10$~kpc. A similar conclusion applies also for the Burkert best 
fit model.

\begin{figure}
 
\vspace{8.5cm}
\includegraphics{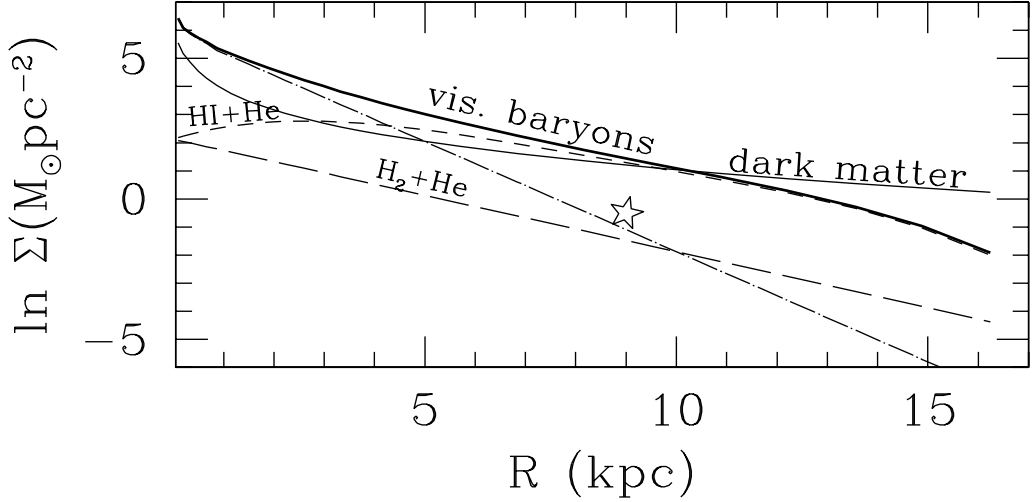}
\caption{The surface density of baryonic and dark matter in M33.
The mass model used for Figure 5$(b)$ has been used here to compute the
stellar contribution from the nucleus and the disk (dot-dashed line)
and the dark matter surface density within 0.5~kpc height from the 
galactic plane (continuous line).
The short dashed line is the HI+He surface density, the long-dashed line 
is the H$_2$+He surface density, the solid heavy line is the sum of the 
two and of the stellar surface density.}
\label{fig:fig7 }
\end{figure}

Using our accurate measurement of the rotational velocity gradient and 
of the surface density of the baryonic and non baryonic components, we 
can check if the onset of massive star formation in M33 is strongly 
related to a global instability of the disk. One of the most popular
tools used to assess the gravitational stability of gaseous disks is
the Toomre stability criterion:

\begin{equation}
 f\Sigma_g < \Sigma_T \equiv {\kappa c_g \over \pi G}
\end{equation}

\noindent
where $\Sigma_g$ is the gas surface mass density, $f$ is the disk thickness 
factor, $\kappa$ is the epicyclic frequency (Toomre 1964), G is the 
gravitational constant, $c_g$ is the gas velocity dispersion. We shall use 
both $f=1$ as for a thin disk and $f=0.7$ to mimic instead a disk with a 
finite thickness (Romeo 1992). The criterion described by equation (7) is 
not at all related to the star formation efficiency but it correctly predicts 
the occurrence of a star formation threshold radius which, in many spiral 
galaxies, has been determined from the drop of the H$\alpha$ surface brightness
(e.g. Martin $\&$ Kennicutt 2001). In M33 Kennicutt (1989) observed such a drop  
at $R_*\simeq 6.4$~kpc. However, since the early work of Kennicutt it 
is well known that the simple Toomre stability criterion fails to account for 
the active star forming region in the disk of low mass spirals such as M33
(Elmegreen 1993, van Zee et al. 1997, Wong $\&$ Blitz 2002). 
An alternative stability criterion includes the shear rate of the disk
and it may be more appropriate for irregular galaxies (Elmegreen 1993,  
Hunter Elmegreen $\&$ Baker 1998). It is based 
on the consideration that making new stars is possible only when there is
enough time for the instability to grow, before shear stretches the
perturbation apart. This corresponds to the following condition:

\begin{equation}
 f\Sigma_g < \Sigma_A \equiv {2.5 A c_g \over \pi G}   
\end{equation}

\noindent
where the Oort constant $A=-0.5Rd\Omega/dR$ is the local shear rate. 
In Figure 8 we display $\Sigma_T$ and  $\Sigma_A$,
using $c_g=6$~km~s$^{-1}$. This value is at the lowest end of the cold gas  
velocity dispersion range estimated in M33 (6-9 km~s$^{-1}$, Warner at al. 1973, 
Thilker 2000). The thin and thick continuous lines in Figure 8 show the mass
surface densities for $f=1$ and for $f=0.7$ respectively. The arrow indicates 
$R_*$. We can see in Figure 8$(a)$ that at $R\simeq R_*$ the azimuthally 
averaged surface gas density for a thin disk falls below the threshold for a 
gravitational instability to develop according to the shear rate criterion.
We shall investigate now if considering additional sources of gravity in the
disk, a modified version of the classical Toomre criterion can still give
correct predictions for the star formation threshold radius.

Since the masses of stars and gas in the M33 disk are comparable, we are 
encouraged to use a stability parameter for a two components fluid. Following
Wang $\&$ Silk (1994) we define $\Sigma_*\equiv \Sigma_s\times c_g/c_s$, 
where $\Sigma_s$ and $c_s$ are the surface density and velocity dispersion 
of the stars in the disk. Using the epicyclic frequency or the shear rate 
of the disk, the stability condition reads:  

\begin{equation}
f(\Sigma_g+\Sigma_*) < \Sigma_T,\Sigma_A
\end{equation}

\begin{figure}
 
\vspace{14.5cm}
\includegraphics{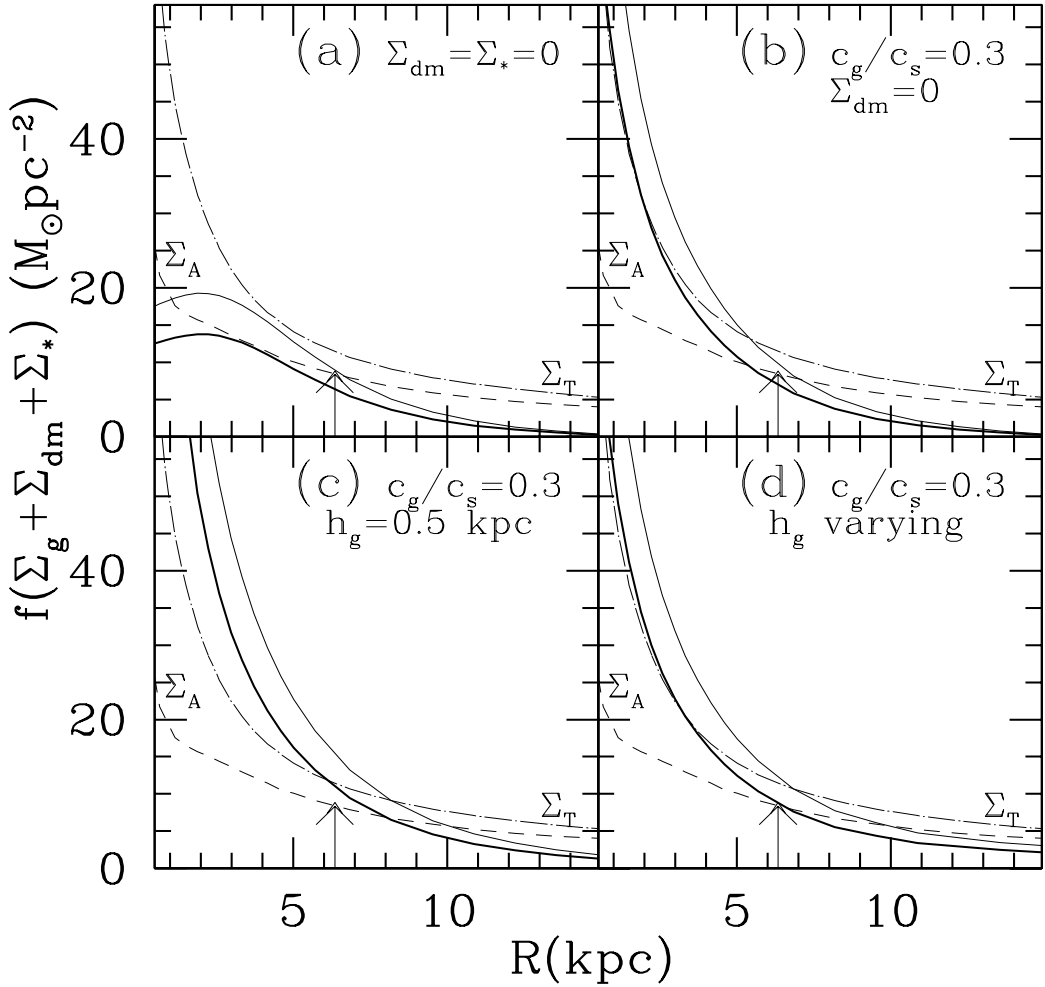}
\caption{The stability of the M33 disk for $c_g=6$~km s$^{-1}$. 
The dashed-dotted line and the dashed line in each panel show 
$\Sigma_T$ and $\Sigma_A$ respectively, as a function of radius. 
An arrow  has been placed at the observational value of $R_*$. 
The thin and thick continuous lines are for $f=1$ and $f=0.7$
respectively. We plot $f\Sigma_g$ in $(a)$  and $f(\Sigma_g+\Sigma_*)$ in 
$(b)$ for $M_d/L=0.7$ and. In $(c)$ we add the dark matter
surface density within a 0.5 kpc layer above and below the galactic plane, 
while in $(d)$  the dark matter is computed within a radially varying 
vertical height of the gas (see text for details).}
\label{fig:fig8 }
\end{figure}

\noindent
Figure 8$(b)$ shows $f(\Sigma_g+\Sigma_*)$, using $M_d/L=0.7$~M$_\odot$/L$_\odot$ 
for deriving $\Sigma_*$ (as for the mass model shown in Fig. 5$(b)$), 
and $c_g/c_s=0.3$.  For $c_g=6$~km~s$^{-1}$ this last ratio implies
$c_s=20$~km~s$^{-1}$, a value observed in the central regions of M33 (Kormendy 
and McClure 1993, Gebhardt et al. 2001). We can see that with a value of 
$M_d/L$ slightly higher than 0.7, the Toomre criterion for a thin disk predicts 
correctly a gravitationally unstable disk for $R<R_*$. If finite thickness effects 
are taken into account or if $c_g$ is higher than 6~km~s$^{-1}$ at $R\simeq R_*$ 
we need the shear rate criterion.

Following Hunter, Elmegreen $\&$ Baker (1998) we consider also the possible 
influence of dark matter gravity. If $\Sigma_{dm}$ is the dark matter surface 
density within the vertical extent of the gas, we can rewrite equation (9) as

\begin{equation}
  f(\Sigma_g+\Sigma_{dm}+\Sigma_*) < \Sigma_T,\Sigma_A
\end{equation} 

\noindent
In Figure 8$(c)$ we show $f(\Sigma_g+\Sigma_{dm}+\Sigma_*)$
assuming a gaseous extent of 0.5~kpc above and below the galactic plane and 
a NFW dark matter profile as used for the rotation curve fit displayed in  
Figure 5$(b)$. In this case finite thickness effects, or equivalently a 
velocity dispersion as high as 9~km~s$^{-1}$, make the disk gravitationally 
unstable, according to the Toomre criterion, exactly for $R<R_*$. This
solution is not unique. If the gas is in hydrostatic equilibrium for example, 
the gas vertical extension and the velocity dispersion are not independent 
variables. An approximation to the scale height of the gas for a galactic disk 
of stars and gas in hydrostatic equilibrium in a dark matter halo is given by

\begin{equation}
h_g={c_g\over \pi G}\Bigl({\Sigma_g+\Sigma_{dm}\over c_g} +
{\Sigma_s\over c_s}\Bigr)^{-1} 
\end{equation}

\noindent
Since $\Sigma_{dm}$ is the dark matter surface density between $+h_g$ and
$-h_g$ the above formula should be solved numerically for determining $h_g$ 
and $\Sigma_{dm}$. In Figure 8$(d)$ we plot $f(\Sigma_g + \Sigma_{dm}+
\Sigma_{*})$, for $\Sigma_{dm}$ between $\pm h_g$ for the same dark matter 
density used in Figure 8$(c)$. We find that $h_g$ varies between 20~pc and 
1~kpc through the radial extent of the disk analyzed in this paper. This model 
reduces the radii of the unstable star forming regions respect to the constant
scale height model displayed in Figure 8$(c)$, making both $\Sigma_A$ and 
$\Sigma_T$ suitable indicators of the growth of instabilities in the disk. 
Which one it is better to use for defining a threshold radius in M33 is therefore 
still unclear since this depends on the radial variations of the involved 
parameters such as sound speed, thickness etc. which are not well determined yet.
It is important to notice however that since the densities of visible baryons 
and dark matter are well constrained at intermediate radii, the intersections
between $\Sigma_A$ or $\Sigma_T$ and $f(\Sigma_g + \Sigma_{dm} + \Sigma_{*})$
do not depend on which mass model we use, as soon as it is compatible with the 
rotational velocity data. The Burkert best fitting mass model, for example, 
gives the same solutions displayed in Figure 8 for a NFW dark matter halo.

\section{Summary}

A full map of the nearby galaxy M33 in the CO J=1-0 transition has given 
additional data for improving the accuracy of the rotation curve of this galaxy 
and the knowledge of the visible baryons distribution. The molecular gas peaks
in the central regions with a surface density of about 7~M$_\odot$ pc$^{-2}$
and declines exponentially with a radial scalelength of about 2.5~kpc down to 
about 0.6~M$_\odot$ pc$^{-2}$ at $R\simeq 6$~kpc. Our sensitivity was not 
sufficient to outline the molecular gas radial distribution further out and in 
particular to detect any possible sudden drop of the azimuthally averaged 
molecular gas surface density. The molecular gas is not
smoothly distributed in the inner disk since there are remarkable mass 
concentrations along the spiral arms. Some wiggles of about 10 km s$^{-1}$ 
are visible in the rotation curves of the two separate halves of the galaxy, 
which are reminiscent of density wave perturbations.
Details of atomic and ionized gas distribution in M33 can be found in 
Hoopes $\&$ Walterbos (2000) and in Thilker, Braun $\&$ Walterbos (2001).
The rotation curve derived using H$\alpha$ data for the innermost point,
the CO data and the HI 21-cm line data, has been used to constrain the visible
baryonic mass and dark matter halo models. The total gas mass (HI+H$_2$+He) 
is $\sim3.2\times 10^9$~M$_\odot$. This is of the same order of the stellar 
disk mass, estimated to be between 3$\times 10^9$ and 6$\times 10^9$ M$_\odot$.
The total visible baryonic mass of M33 can therefore be as high as 
$10^{10}$~M$_\odot$. The dark halo mass out to the last measured point
($17$~kpc from the center) is $\sim5\times10^{10}$~M$_\odot$. 

Both the non-singular isothermal halo and the Burkert halo model, give good fits 
to the M33 rotation curve when the flat density core is extended. The pure 
$R^{-2}$ or $R^{-3}$ outer scaling law applies only beyond the outermost observed 
radius. Dark matter density profiles with an inner $R^{-1.5}$ cusp   
(Moore et al. 1998) are too steep to be consistent with the rotation of the 
central regions of M33 and are ruled out. Halo models with milder central cusps,
such as the NFW radial profile, fit well all the observed rotational velocities.  
The mass of the virialized NFW dark halo for this galaxy is 
$\ge 5\times 10^{11}$~M$_\odot$ and the halo size comparable to the distance 
between M33 and its bright companion M31 ($\sim 180$~kpc). The baryon fraction 
in M33 is then $\le 0.02$. The required concentrations $C_\Delta$ are below the 
scaling laws found by numerical simulations of structures formation in a 
standard Cold Dark Matter scenario ($\Omega_0=0.3$, $\Lambda_0=0.7$, 
$\sigma_8$=0.9) and favor cosmological models which predicts less concentrated 
dark halos (Eke, Navarro $\&$ Steinmetz 2001, Zentner $\&$ Bullock 2002, van 
den Bosch, Mo $\&$ Yang 2003). We have shown however that if  
an associated scatter in the numerical simulation results is considered 
(Wechsler et al. 2002), the M33 dark halo is still compatible with 
galaxy formation models in a standard $\Lambda$CDM cosmology.
The finding that dark matter halos are consistent with constant density 
core models or with central cusps less steep than $R^{-1.5}$, is known as
the cusp-core degeneracy (van den Bosch $\&$ Swaters 2001). It is
remarkable that this degeneracy holds also in M33 which is a galaxy with
small uncertainties about its distance, its gaseous and stellar content,  
and with circular velocities sampled over a wide range of radii: from 
galactocentric distances of a few parsecs out to 13 disk scalelengths. 
For this galaxy it seems that the degeneracy holds mostly because of a  
nuclear stellar component, which regulates the dynamics of the innermost 
0.5~kpc region.  This extended `stellar nucleus', with mass between 
$3\times 10^7$ and $8\times 10^8$~M$_\odot$, is not only dynamically 
required by the rotational data but it is also evident from photometric data
(Bothun 1992, Minniti, Olszewski $\&$ Rieke 1993, Regan $\&$ Vogel 1994). 
We have considered a core collapse model and a de Vaucouleurs $R^{1/4}$ law  
for the nuclear stellar density but additional data on the kinematics 
and light distribution of the central few hundreds parsec region are needed 
to constrain better the mass distribution there. 

Having a very accurate rotation curve and a good estimate of the gas, stellar
and dark matter surface density in the M33 disk, we have addressed the
question on the possible coincidence of the active star formation region in 
this galaxy with a gravitationally unstable region. Azimuthal averages 
of the observed gas surface densities are above the critical densities in the 
star forming region of M33 only if one considers the minimum acceptable value 
of the gas velocity dispersion  and a stability criterion based on the shear 
rate of the disk. However, the simple Toomre condition predicts a star formation
threshold radius in agreement with the observed drop in the H$\alpha$ surface
brightness if the additional compression due to the stellar disk or to the dark
matter surface density is considered.
               
I acknowledge Steve Schneider and Mark Heyer for having given me the opportunity
to use the CO data prior to publication. I am also very thankful to the
anonymous referee who made several excellent suggestions for improving the 
quality of the work presented in the original manuscript.

\label{lastpage}

\end{document}